\documentclass[twocolumn,prl]{revtex4}

\usepackage{graphicx}

\begin{document}


\title{Braid Topologies for Quantum Computation}
\author{N.~E. Bonesteel, L. Hormozi and G. Zikos}
\affiliation{Department of Physics and National High Magnetic
Field Laboratory, Florida State University, Tallahassee, Florida
32310}
\author{S.~H. Simon}
\affiliation{Bell Laboratories, Lucent Technologies, Murray Hill,
New Jersey 07974}

\begin{abstract}
In topological quantum computation, quantum information is stored
in states which are intrinsically protected from decoherence, and
quantum gates are carried out by dragging particle-like
excitations (quasiparticles) around one another in two space
dimensions. The resulting quasiparticle trajectories define
world-lines in three dimensional space-time, and the corresponding
quantum gates depend only on the topology of the braids formed by
these world-lines. We show how to find braids that yield a
universal set of quantum gates for qubits encoded using a specific
kind of quasiparticle which is particularly promising for
experimental realization.
\end{abstract}


\maketitle

A quantum computer must be capable of manipulating quantum
information while simultaneously protecting it from error and loss
of quantum coherence due to coupling to the environment.
Topological quantum computation (TQC) \cite{kitaev,freedman}
offers a particularly elegant way to achieve this using
quasiparticles which obey nonabelian statistics
\cite{frohlich,moore}. These quasiparticles, which are expected to
arise in a variety of two-dimensional quantum many-body systems
\cite{kitaev,moore,read,slingerland,cooper,duan,doucot,nayak,fendley},
have the property that the usual phase factors of $\pm 1$
associated with the exchange of identical bosons or fermions are
replaced by noncommuting (nonabelian) matrices that depend only on
the topology of the space-time paths (braids) used to effect the
exchange. The matrices act on a degenerate Hilbert space whose
dimensionality is exponentially large in the number of
quasiparticles and whose states have an intrinsic immunity to
decoherence because they cannot be distinguished by local
measurements, provided the quasiparticles are kept sufficiently
far apart.

In TQC  this protected Hilbert space is used to store quantum
information, and quantum gates are carried out by adiabatically
braiding quasiparticles around each other \cite{kitaev,freedman}.
Because the resulting quantum gates depend purely on the topology
of the braids, errors only occur when quasiparticles form
``unintentional" braids. This can happen if a
quasiparticle-quasihole pair is thermally created, the pair
separates, wanders around other quasiparticles, and then
recombines in a topologically nontrivial way. However, such
processes are exponentially unlikely at low enough temperature.
This built in protection from error and decoherence is an
appealing feature of TQC which may compensate for the extreme
technical challenges that will have to be overcome to realize it.

It has been shown that several different kinds of nonabelian
quasiparticles can be used for TQC
\cite{kitaev,freedman,ogburn,mochon,preskill}. Here we focus on
what is arguably the simplest of these --- Fibonacci anyons
\cite{preskill}. These quasiparticles each possess a ``q-deformed"
spin quantum number (q-spin) of 1, the properties of which are
described by a mathematical structure known as a quantum group
\cite{fuchs}. As with ordinary spin there are specific rules for
combining q-spin. For Fibonacci anyons these ``fusion" rules state
that when two q-spin 1 objects are combined, the total q-spin can
be either 0 or 1; and when a q-spin 0 object is combined with a
q-spin $s$ object, where $s =$ 0 or 1, the total q-spin is $s$
\cite{m}. Remarkably, as shown in \cite{preskill}, these fusion
rules fix the structure of the relevant quantum group, uniquely
determining the quantum operations produced by braiding q-spins
around one another up to an overall abelian phase which is
irrelevant for TQC.

One reason for focusing on Fibonacci anyons is that they are
thought to exist in an experimentally observed fractional quantum
Hall state \cite{pan,twelvefifths}. It may also be possible to
realize them in rotating Bose condensates \cite{cooper} and
quantum spin systems \cite{nayak,fendley}. Strictly speaking, the
quantum group realized in some of these systems, and considered
for TQC in \cite{freedman}, also includes q-spins of $\frac{1}{2}$
and $\frac{3}{2}$; however, due to a symmetry of this quantum
group \cite{slingerland}, the braiding properties of q-spin
$\frac{1}{2}$ quasiparticles are equivalent to those with q-spin 1
and the braid topologies we find below can be used in either case.

The fusion rules for Fibonacci anyons imply the Hilbert space of
two quasiparticles is two dimensional --- with basis states
$|(\bullet,\bullet)_0\rangle$ and $|(\bullet,\bullet)_1\rangle$.
Here the notation $(\bullet,\bullet)_a$ represents two
quasiparticles with total q-spin $a$. When a third quasiparticle
is added, the Hilbert space is three dimensional, and is spanned
by the states $|((\bullet,\bullet)_0,\bullet)_1\rangle$,
$|((\bullet,\bullet)_1,\bullet)_1\rangle$ and
$|((\bullet,\bullet)_1,\bullet)_0\rangle$. The general result is
that the dimensionality of an $N$-quasiparticle state is the
$(N+1)^{\rm st}$ Fibonacci number. To use this Hilbert space for
quantum computation we follow Freedman et al. \cite{freedman}, and
encode qubits into triplets of quasiparticles with total q-spin 1,
taking the logical qubit states to be $|0_L\rangle =
|((\bullet,\bullet)_0,\bullet)_1\rangle$ and $|1_L\rangle =
|((\bullet,\bullet)_1,\bullet)_1\rangle$.  The remaining state
with total q-spin 0 is then a noncomputational state, $|NC\rangle
= |((\bullet,\bullet)_1,\bullet)_0\rangle$.  This encoding,
illustrated in Fig.~1, can be viewed as a q-deformed version of
the three-spin qubit encoding proposed for exchange-only quantum
computation \cite{exchangeonly}. As in that case, qubits can be
measured by determining the q-spin of the two leftmost
quasiparticles, either by performing local measurements once the
quasiparticles are moved close together \cite{kitaev}, or possibly
by performing interference experiments \cite{overbosch,dassarma}.
Similar schemes can be used for initialization. The price for
introducing this encoding is that care must now be taken to
minimize transitions to noncomputational states, known as leakage
errors, when carrying out computations.

\begin{figure}[t]
\centerline{\includegraphics[scale=.25]{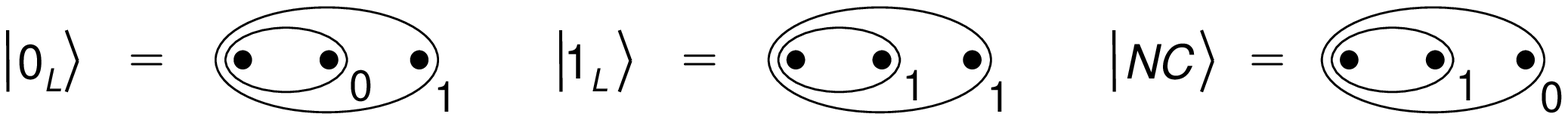}} \caption{Basis
states for the three dimensional Hilbert space of three
quasiparticles and qubit encoding. The ovals enclose groups of
quasiparticles in q-spin eigenstates labeled by the corresponding
eigenvalues. The states $|0_L\rangle$ and $|1_L\rangle$ (denoted,
respectively, $|((\bullet,\bullet)_0,\bullet)_1\rangle$ and
$|((\bullet,\bullet)_1,\bullet)_1\rangle$ in the text) span the
computational qubit space, while the state $|NC\rangle$ (denoted
$|((\bullet,\bullet)_1,\bullet)_0\rangle$ in the text) is a
noncomputational state.}
\end{figure}

Figure 2(a) shows elementary braiding operations for three
quasiparticles together with the matrices which describe the
transitions they induce in the Hilbert space illustrated in Fig.~1
\cite{freedman,slingerland,preskill}. Any three-quasiparticle
braid can be constructed out of these elementary operations and
their inverses.  The corresponding transition matrix can then be
computed by simply multiplying the appropriate matrices as shown
in Fig.~2(b). The upper 2$\times$2 blocks of these matrices act on
the computational qubit space, and the lower right element is a
phase which multiplies $|NC\rangle$. This block diagonal form
illustrates that if a group of quasiparticles is in a q-spin
eigenstate then braiding of quasiparticles within this group does
not lead to transitions out of this eigenstate.  It follows that
single qubit gates performed by braiding quasiparticles within a
qubit will not lead to leakage error.

\begin{figure}[t]
\includegraphics[scale=.32]{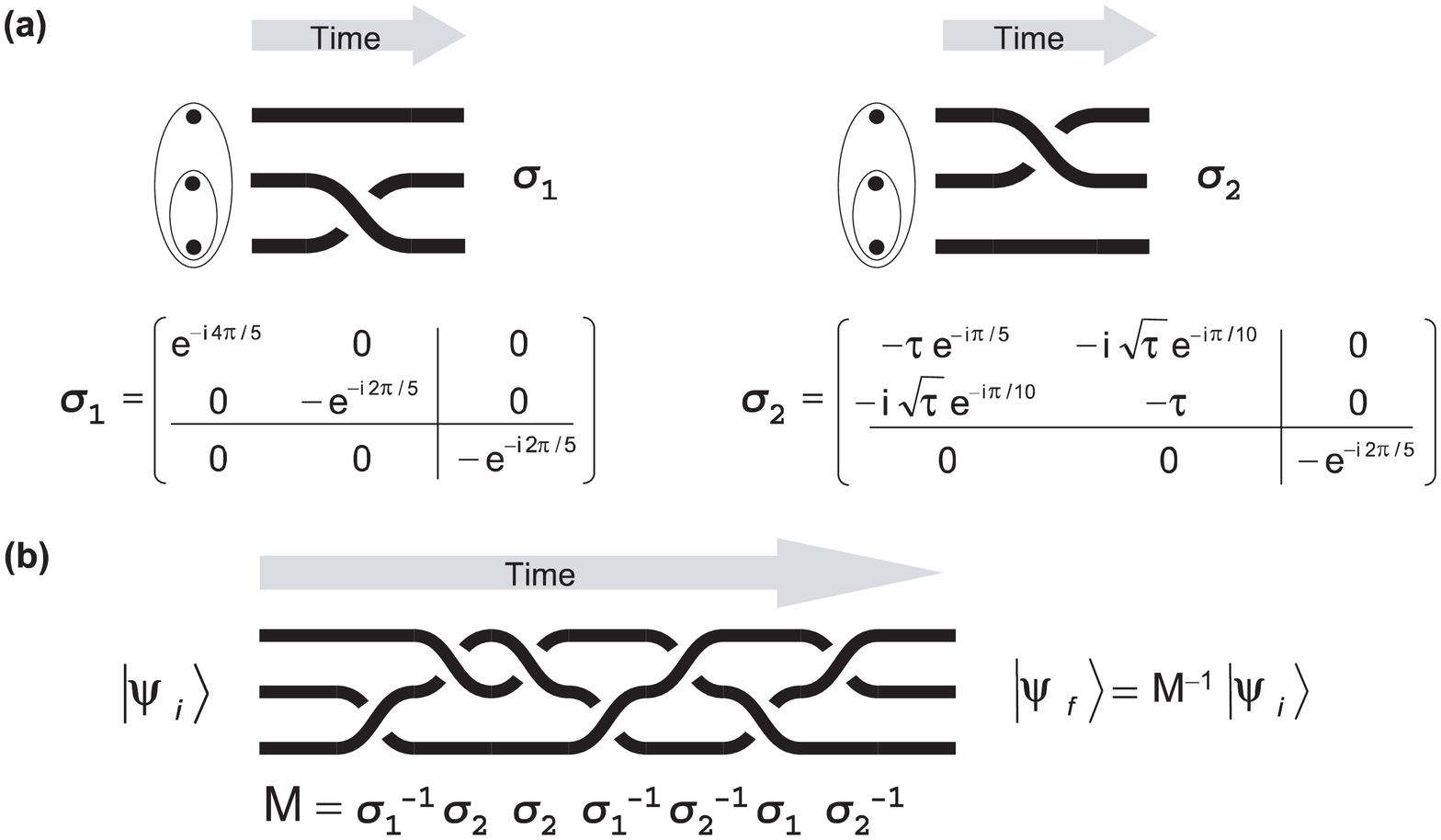}
\caption{(a) Elementary three-quasiparticle braids. The pictures
represent quasiparticle world-lines in 2+1 dimensional space-time,
with time flowing from left to right. The matrices $\sigma_1$ and
$\sigma_2$ are the transition matrices produced by these elementary
braids which act on the three dimensional Hilbert space shown in
Fig.~1. Here $\tau = (\sqrt{5}-1)/2$ is the inverse of the golden
mean. The upper 2$\times$2 blocks of these matrices act on the
computational qubit space (total q-spin 1) and are used to perform
single qubit rotations, while the lower right element is a phase
which multiplies $|NC\rangle$. (b) A general three-quasiparticle
braid and the corresponding matrix expression for the transition
matrix it produces.  Here $|\psi_i\rangle$ is the initial state and
$|\psi_f\rangle$ the final state after braiding. Note that these
(and subsequent) figures only represent the topology of the braid.
In any actual implementation quasiparticles will have to be kept
sufficiently far apart to keep from lifting the topological
degeneracy.}\ \vskip -.25in
\end{figure}

To find braids which perform a given single qubit gate we first
carry out a brute force search of three-quasiparticle braids with up
to 46 interchanges. This exhaustive search typically yields braids
approximating the desired target gate to within a distance of
$\epsilon \sim 1-2 \times 10^{-3}$ (here we define distance between
gates using the operator norm -- see Fig.~3 for a definition). If
more accuracy is required, brute force searching becomes
exponentially more difficult and rapidly becomes unfeasible.
Fortunately, a powerful theorem due to Solovay and Kitaev
\cite{kitaevbook,nielsenbook} guarantees that given a set of gates
generated by finite braids which is sufficiently dense in the space
of all gates, (easily generated for three quasiparticles), braids
approximating arbitrary single-qubit gates to any required accuracy
can be found efficiently, with the length of the braid growing as
$\sim |\log \epsilon|^c$ where $c \simeq 4$.

We now turn to the significantly more difficult problem of finding
braids which approximate a desired two-qubit gate. In this case
there are six quasiparticles, and the Hilbert space is thirteen
dimensional. The elementary braid matrices acting on this space are
again block diagonal,  with 5$\times$5 (total q-spin 0) and
8$\times$8 (total q-spin 1) blocks \cite{fusionspace}.  It is known
that braiding these six quasiparticles generates a set of unitary
operations which is dense in the space of all such block diagonal
operations \cite{freedman,preskill}, and the Solovay-Kitaev theorem
again guarantees one can in principle construct braids to
approximate any desired operation of this form \cite{kitaevbook}.
However, unlike the single qubit case, actual implementation of this
procedure is problematic. The space of unitary operations for six
quasiparticles is parameterized by 87 continuous parameters, as
opposed to 3 for the three quasiparticle case, and searching for
braids which approximate a desired quantum gate in this high
dimensional space is an extremely difficult numerical problem. To
circumvent this difficulty, we have found constructions for a
particular class of two-qubit gates (controlled rotation gates)
which only require finding a finite number of three-quasiparticle
braids. The resulting reduction of the dimensionality of the search
space from 87 to 3 makes it possible for the first time to compile
accurate braids for a class of two-qubit gates which can be
systematically improved using the Solovay-Kitaev theorem.

\begin{figure}[t]
\includegraphics[scale=.35]{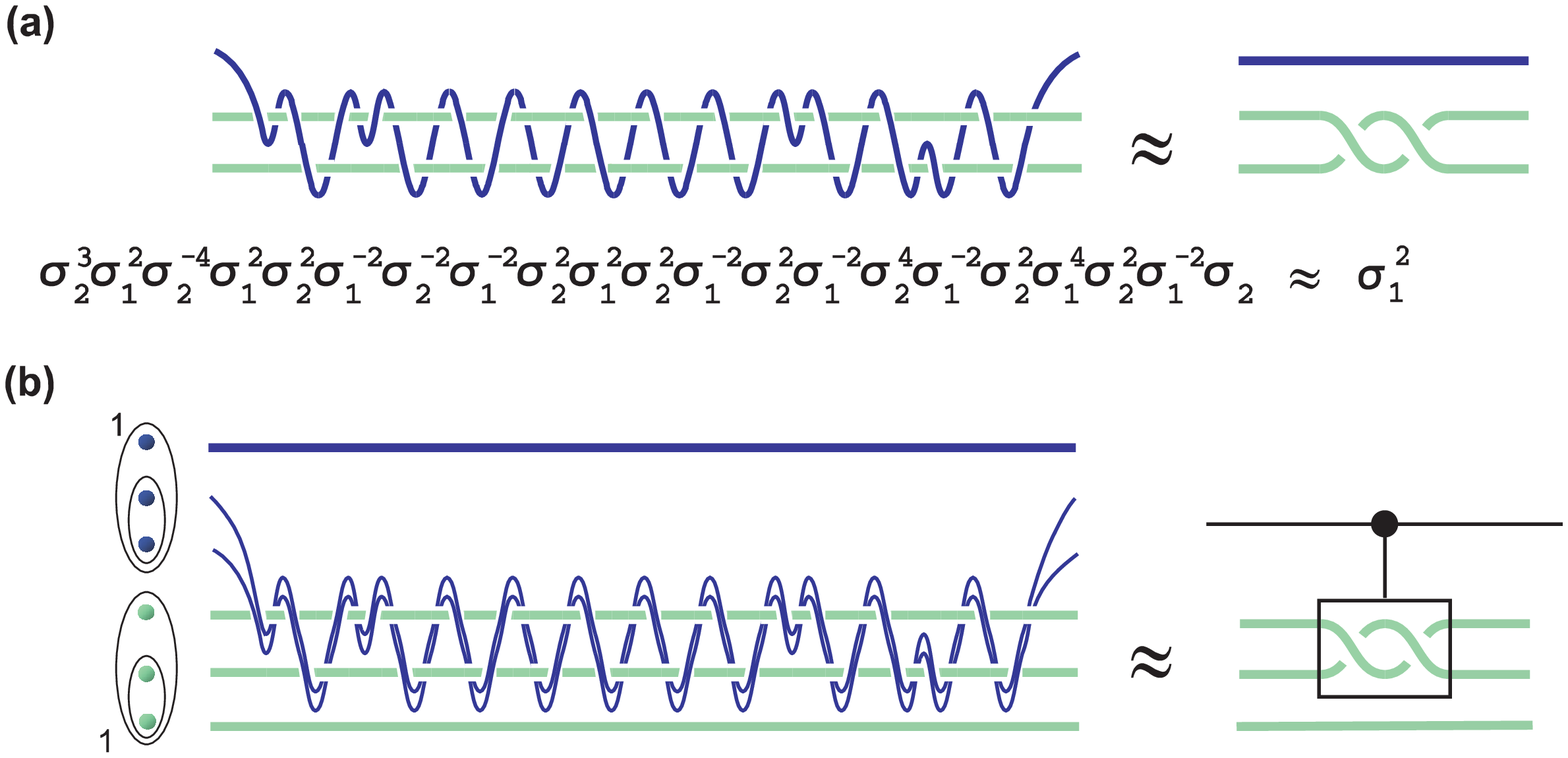}
\caption{(Color online). (a) A three-quasiparticle braid in which
one quasiparticle is woven around two static quasiparticles and
returns to its original position (left), and yields approximately
the same transition matrix as braiding the two stationary
quasiparticles around each other twice (right). The corresponding
matrix equation is also shown. To characterize the accuracy of this
approximation we define the distance between two matrices, $U$ and
$V$, to be $\epsilon = ||U-V||$, where $||O||$ is the operator norm
of $O$ equal to the square-root of the highest eigenvalue of
$O^\dagger O$. The distance between the matrices resulting from the
actual braiding (left) and the desired effective braiding (right) is
$\epsilon \simeq$ 2.3$\times$10$^{-3}$. ({\bf b}) A two-qubit braid
constructed by weaving a pair of quasiparticles from the control
qubit (top) through the target qubit (bottom) using the weaving
pattern from (a). The result of this operation is to effectively
braid the upper two quasiparticles of the target qubit around each
other twice if the control qubit is in the state $|1_L\rangle$, and
otherwise do nothing.  This is an entangling two-qubit gate which
can be used for universal quantum computation. Since all effective
braiding takes place within the target qubit, any leakage error is
due to the approximate nature of the weave shown in (a). By
systematically improving this weave using the Solovay-Kitaev
construction, leakage error can be reduced to whatever level is
required for a given computation.}\ \vskip -.25in
\end{figure}

Our constructions are based on two essential ideas.  First, we
{\it weave} a pair of quasiparticles (the control pair) from one
qubit (the control qubit) through the quasiparticles forming the
second qubit (the target qubit). By weaving we mean that the
target quasiparticles remain fixed while the control pair is moved
around them as an immutable group (see, for example, Figs.~3(b)
and 4(c)). If the q-spin of the control pair is 0 the result of
this operation is the identity. However, if the q-spin of the
control pair is 1, a transition is induced. If we choose the
control pair to consist of the two quasiparticles whose total
q-spin determines the state of the control qubit, this
construction automatically yields a controlled (conditional)
operation. Second, we deliberately weave the control pair through
only two target quasiparticles at a time. Since the only
nontrivial case is when the control pair has q-spin 1, and is thus
equivalent to a single quasiparticle, this reduces the problem of
constructing two-qubit gates to that of finding a finite number of
specific three-quasiparticle braids.

Figure 3(a) shows a three-quasiparticle braid in which one
quasiparticle is woven through the other two and then returns to its
original position. The resulting unitary operation approximates that
of simply braiding the two static quasiparticles around each other
twice to a distance of $\epsilon \simeq 2.3 \times 10^{-3}$. Similar
weaves can be found which approximate any even number, $2m$, of
windings of the static quasiparticles. Figure 3(b) shows a two-qubit
braid in which the pattern from Fig.~3(a) is used to weave the
control pair through the target qubit. If the control qubit is in
the state $|0_L\rangle$ this weave does nothing, but if it is in the
state $|1_L\rangle$ the effect is equivalent to braiding two
quasiparticles within the target qubit. Thus, in the limit $\epsilon
\rightarrow 0$, this {\it effective} braiding is all within a qubit
and there are no leakage errors. The resulting two-qubit gate is a
controlled rotation of the target qubit through an angle of
$6m\pi/5$, which, together with single qubit rotations, provides a
universal set of gates for quantum computation provided $m$ is not
divisible by 5 \cite{bremner}. Carrying out one iteration of the
Solovay-Kitaev construction \cite{kitaevbook,nielsenbook} on this
weave using the procedure outlined in \cite{harrow} reduces
$\epsilon$ by a factor of $\sim 20$ at the expense of a factor of 5
increase in length. Subsequent iterations can be used to achieve any
desired accuracy.

\begin{figure*}[t]
\centerline{\includegraphics[scale=.5]{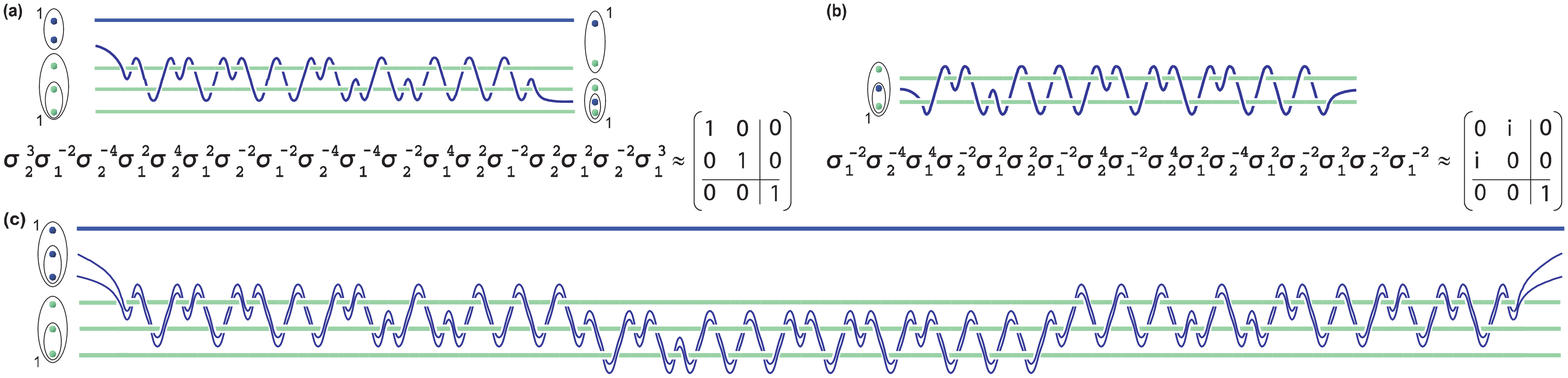}}
\caption{(Color online). (a) An injection weave for which the
product of elementary braiding matrices, also shown, approximates
the identity to a distance of $\epsilon \simeq$
1.5$\times$10$^{-3}$. This weave injects a quasiparticle (or any
q-spin 1 object) into the target qubit without changing any of the
underlying q-spin quantum numbers. (b) A weaving pattern which
approximates a NOT gate to a distance of $\epsilon \simeq$
8.5$\times$10$^{-4}$. (c) A controlled-NOT gate constructed using
the weaves shown in (a) and (b) to inject the control pair into the
target qubit, perform a NOT operation on the injected target qubit,
and then eject the control pair from the target qubit back into the
control qubit. The distance between the gate produced by this braid
acting on the computational two-qubit space and an exact
controlled-NOT gate is $\epsilon \simeq$ 1.8 $\times$10$^{-3}$ and
$\epsilon \simeq$ 1.2 $\times$10$^{-3}$ when the total q-spin of the
six quasiparticles is 0 and 1, respectively. Again, the weaves shown
in (a) and (b) can be made as accurate as necessary using the
Solovay-Kitaev theorem, thereby improving the controlled-NOT gate to
any desired accuracy. By replacing the central NOT weave, arbitrary
controlled rotation gates can be constructed using this
procedure.}\vskip -.1in
\end{figure*}

A similar construction can be used to carry out arbitrary
controlled-rotation gates. Figure 4(a) shows a braid in which one
quasiparticle is again woven through two static quasiparticles, but
this time does not return to its original position. The unitary
transformation produced by this weave approximates the identity
operation to a distance of $\epsilon \simeq 1.5 \times$10$^{-3}$,
where, as above, the accuracy of this approximation can be
systematically improved by the Solovay-Kitaev theorem.  In the limit
$\epsilon \rightarrow 0$, the effect of this weave is to permute the
three quasiparticles involved without changing any of the underlying
q-spin quantum numbers, as shown in the figure. It can therefore be
used to safely {\it inject} a quasiparticle, or any object with
q-spin 1, into a qubit.  Figure 4(b) then shows a weave which
performs an approximate NOT gate on the target qubit.  These two
weaves are used to construct the two-qubit braid shown in Fig.~4(c).
In this braid, the control pair is first injected into the target
qubit using the ``injection weave." When the control pair has q-spin
1 the state of the modified target qubit is unchanged after
injection -- the only effect is that one of the target
quasiparticles has been replaced by the control pair. A NOT
operation is then performed on the injected target qubit by weaving
the control pair inside the target using the pattern from Fig.~4(b).
In the limit of an exact injection weave this braiding is all within
a q-spin eigenstate and there are no leakage errors. Finally the
control pair is {\it ejected} from the target using the inverse of
the injection weave, thereby returning the control qubit to its
original state. As before, if the control qubit is in the state
$|0_L\rangle$ the result is the identity. However, if the control
qubit is in the state $|1_L\rangle$, a NOT gate is performed on the
target qubit. This construction therefore produces a controlled-NOT
gate, up to single qubit rotations \cite{notnot}. Because a weave
producing any single-qubit rotation can be used instead of the NOT
weave shown in Fig.~4(b) this construction can be used to produce an
arbitrary controlled rotation gate.

We thank D.P. DiVincenzo and D. Stepanenko for useful discussions.
N.E.B. and L.H. are supported by US DOE grant DE-FG02-97ER45639.


\begin{references}
\bibitem{twelvefifths}  A fractional quantum Hall state with
Landau level filling fraction $\nu = 12/5$ has been observed
experimentally \cite{pan} and may be described by the $k=3$
``parafermion" state proposed in \cite{read}.  The quantum group
description of the quasiparticle excitations of this state,
developed in \cite{slingerland}, is the same (up to an irrelevant
abelian phase) as that considered here.
\bibitem{fusionspace} {The existence of these blocks reflects the
fact that for multiple qubits there is, in addition to the
computational qubit space, a `fusion' space associated with applying
the fusion rules to the q-spin 1 quantum numbers of each qubit. To
avoid errors due to entangling this fusion space with the
computational space, the action of all two-qubit gates must be
independent of the total q-spin of the two qubits involved.  By
construction, the gates presented here satisfy this requirement.}
\bibitem{notnot} Single qubit rotations
of the control qubit are required to eliminate the phase factor of
$i$ which appears in the NOT gate shown in Fig.~4(b).
\end{references}
\end{document}